\documentclass[a4paper,12pt]{article}
\pdfoutput=1 

\usepackage{jheppub} 

\usepackage[T1]{fontenc} 

\usepackage{wasysym}
\usepackage[usenames,dvipsnames]{xcolor}

\usepackage{amsfonts}
\usepackage[english]{babel}
\usepackage{ucs}
\usepackage[utf8x]{inputenc}
\usepackage{lmodern}
\usepackage[T1]{fontenc}
\usepackage{graphicx,amssymb,amsmath,bm}
\usepackage{slashed,epstopdf}
\usepackage{subfig}
\usepackage{tikz}
\usepackage{mathtools}

\definecolor{beaver}{rgb}{0.62, 0.51, 0.44}

\definecolor{auburn}{rgb}{0.43, 0.21, 0.1}

\definecolor{amethyst}{rgb}{0.6, 0.4, 0.8}
\newcommand{\amethyst}[1]{{\color{amethyst}}}
\newcommand{\beq}{\begin{eqnarray}}
\newcommand{\eeq}{\end{eqnarray}}

\def\Dsl{\,\raise.15ex\hbox{/}\mkern-12.5mu D}

\newcommand{\xddots}

\usepackage [english]{babel}
\usepackage{amsfonts}
\usepackage{amssymb}
\usepackage{epsf}
\usepackage{latexsym,amssymb,euscript}
\usepackage{graphicx}
\usepackage{amsmath}
\input epsf

\def\@keyword{}
\newcommand{\keyword}[1]{\gdef\@keyword{#1}}
\def\@pacs{}
\newcommand{\PACS}[1]{\gdef\@pacs{#1}}

\date{}
\title{
Asymptotic freedom and  higher derivative gauge theories}
\author[a]{M.  Asorey}
\author[a]{F.  Falceto} 
\affiliation[a]{Centro de Astropart\'{\i}culas y F\'{\i}sica de Altas Energ\'{\i}as, Departamento de F\'{\i}sica Te\'orica\break
Universidad de Zaragoza, C/ Pedro Cerbuna 12, E-50009 Zaragoza, Spain}
\author[b]{L.  Rachwa\l{}}
\affiliation[b]{\small Departamento de F\'{\i}sica -- ICE, Universidade Federal de Juiz de Fora, 
Rua Jos\'e Louren\c co Kelmer, Campus Universit\'ario, Juiz de Fora, 36036-900, MG, Brazil }

\emailAdd{asorey@unizar.es}
\emailAdd{falceto@unizar.com}
\emailAdd{grzerach@gmail.com}

\abstract
{The ultraviolet completion of gauge theories by higher derivative terms can dramatically  change their
behavior at high energies.
The requirement of asymptotic freedom  imposes very stringent constraints that
 are only satisfied  by a small  family of higher derivative theories.
 If the number of derivatives is large enough ($n > 4$)  the theory is strongly interacting  both at extreme
infrared and ultraviolet regimes whereas it 
 remains asymptotically free  for a low number of extra derivatives ($n \leqslant 4$).    
In all cases   the theory improves its ultraviolet behavior
 leading in some cases to ultraviolet finite theories with vanishing $\beta$-function. The usual consistency problems 
 associated to the presence of extra ghosts in higher derivative theories may not harm asymptotically free theories because in that case the effective masses of such ghosts are running to infinity in the ultraviolet limit. }

\makeatletter
\gdef\@fpheader{}
\makeatother
\begin{document}
\notoc
\maketitle
\vfill

\section{Introduction}

Field theories with higher derivatives were first considered as covariant ultraviolet regularizations
of gauge theories \cite{FS}-\cite{LZ}. However, in  the last years there is a renewed interest in these theories
 mainly due to the rediscovery that they provide a renormalizable field theoretical framework
 for the quantization of gravity \cite{Stelle,Stelle2}. Field theories with higher derivatives can be also  considered
 as very efficient effective theories in strongly correlated regimes of standard gauge field theories and
extended gravity theories in inflationary scenarios \cite{Starobinsky, FT}.
From a fundamental viewpoint  higher derivatives theories were disregarded in the
past  due to the fact that they face pathological  behaviors concerning causality and unitarity principles
\cite{alsh,arsh}.
However, such a  behavior is not a problem in some higher derivative theories 
such as  Lee-Wick theories \cite{leewick}--\cite{AP2} and theories with    ghost condensation phenomena.
We shall show that the small family of consistent gauge theories may be extended to 
asymptotically free gauge theories.

These theories might provide ultraviolet completions of the Standard Model.
Higher derivative theories of spin-0 and spin-1/2  fields with local interactions are finite.
However,  in general relativity and non-abelian gauge theories  one cannot get rid of
one-loop ultraviolet (UV) divergences that require renormalization \cite{FS}-\cite{LZ}.
We shall show that there are some very special higher derivative theories that do not require
renormalization because all UV divergences cancel out. These finite theories
do not have UV singularities even classically, which is particularly interesting
in gravity theories because they lead to black holes without hidden singularities
and  evolutions of the Universe without Big Bang    singularity    \cite{BGKM, BCKM}.
However, this does not exclude that some of those theories can present infrared (IR) divergences.

In this paper,  we analyze from this perspective the ultraviolet behavior
of higher derivative  gauge theories  
and, in particular, the asymptotic freedom behavior at extremely high energies. We also
analyze the emergence of IR divergences and how the renormalization
group (RG) can smoothly interpolate between these two asymptotic regimes and solve the
consistency problems associated to the appearance of extra ghost fields.


\section{Higher derivative    gauge    theories}

   In the Euclidean formalism    a    pure gauge    theory with higher derivatives can be defined by the following action
\begin{equation}
S=\frac{1}{4  g^2}\int d^{4}x\, F_{\mu\nu}^{a} F^{\mu\nu}{}^{a}         + 
\frac{1}{4  g^2{\Lambda^{2n}}}\int d^{4}x\, F_{\mu\nu}^{a}\, {\Delta ^{n}}\,F^{\mu\nu}{}^{a}\,,
\label{hd}
\end{equation}
where   the operator $\Delta$ defined by
\begin{equation}
\Delta_{\mu\, a}^{\nu\, b}=-D^{2} \delta^{\nu}_\mu\delta^{b}_a +2f^b{}_{ca}
\, F_{\mu}{}^{\nu c}
\label{HL}
\end{equation}
is the Hodge-covariant Laplacian operator $\Delta=d_A^\ast d_A+ d_A d_A^\ast$ acting on the Lie algebra \,{$\mathfrak{g}{\rm-valued}$}  2-forms $F$ of the gauge field strength $F_{\mu\nu}$ 
and
$D_\mu$ is the    gauge-covariant derivative.    The gauge group is denoted by $G$ here and its corresponding algebra by $\mathfrak{g}$. The gauge-covariant exterior differential on forms we denote by $d_A$ and  $d_A^\ast$ is its Hodge-dual.

These new theories are not conformally invariant,    in contrast with standard Yang-Mills  theory in $d=4$ spacetime dimensions.    But they
preserve     all  solutions of the 
Yang-Mills vacuum equation of motion  $d_A^\ast  F =0$, including standard instantons ($F = \pm \ast F $) for Euclidean theory,    as exact solutions of their equations of motion.

Although one-loop corrections around such solutions are different than that of ordinary Yang-Mills  theories
and might have a softer UV behavior, they share the same
pathological IR  behavior,  which usually spoils the instanton physical implications.

 For $n\geqslant 2$ the new theories (\ref{hd}) are superrenormalizable. They only  have one-loop UV divergences \cite{FS,LZ}. Higher order contributions become finite once the one-loop divergences are renormalized \cite{af1, af2}.

The calculation of UV divergences can be performed in  $\alpha$-gauge    by adding the following gauge-fixing functional to the action (\ref{hd})
\begin{equation}
S_\alpha=\frac{\alpha}{2  g^2 \Lambda^{2n}}\int d^4x\, \partial^\mu A^a_\mu (-\partial^\sigma \partial_\sigma)^{n}\partial^\nu A^a_\nu
\end{equation}
and by using dimensional regularization with $\epsilon=4-d$,    where $d$ is the regularized dimension of Euclidean space.    Returning to Minkowski spacetime formalism, the  result
for  the UV-divergent contribution of the two-point gluonic function coming from one-loop vacuum polarization
diagrams is  {(see Appendix)}
\begin{equation}
\Gamma_{\mu\nu}^{ab}(p)=   -
c_n{
{\frac{C_{2}(G)}{ 16\pi^{2}\epsilon}}} i\delta^{ab}\left(p^{2}\eta_{\mu\nu}-p_{\mu}p_{\nu}\right)
\label{Pidefinition}
\end{equation}
with
\begin{equation}c_0= {\alpha}{} -\frac{13}{3},\ \ \
  c_1=- \frac{43}{3}\  \ \  \hbox{and}\ \ \  c_n= 5n^{2}-23n+\frac{29}{3}
  , \, \ \ \hbox{for} \ \ \, n\geqslant 2,  
\  \label{ccoeff}    \end{equation}
which agrees for $n\geqslant 2$ with the results of Asorey-Falceto \cite{af1, af2} and for $n=1$ with those of Babelon-Namazie \cite{babelon}.    In the radiative correction to the two-point function (\ref{Pidefinition}),  $p$ is the momentum of the  incoming gluon,
 $\eta_{\mu\nu}$ denotes the  Minkowski metric
of flat spacetime
 and $C_2(G)$  is the quadratic Casimir  operator of the gauge group $G$.
Notice the independence of $c_n$ on the gauge fixing parameter $\alpha$ for $n\geqslant 1$, because in that case all $\alpha$-dependent terms are finite at any loop order ({see Appendix for a detailed explanation}). It can be shown that in such a case the divergent contributions to 3-point and 4-point
gluonic functions preserve gauge invariance, and
thus, all UV one-loop divergences can be removed  by a simple counterterm
\begin{equation}
S_{\rm count}= c_n\frac{C_{2}(G) }{ 128\pi^{2}} \left(\frac2{\epsilon} +\log {\displaystyle \frac{\Lambda^2_{\rm QCD}}{\Lambda^2} }\right)\, F_{\mu\nu}^{a}   F^{\mu\nu}{}^{a}  ,
\end{equation}
 where the $\Lambda_{\rm QCD}$ scale has been introduced by a renormalization prescription, which adds a finite counterterm  to the minimal renormalization scheme to recover in the IR the renormalized two-point function of  standard  QCD towards  our higher derivative theory (\ref{hd}) tends to.

In the case $n=1$, there are still some two-loop divergent contributions which require additional
 renormalization. In the case $n=0$, besides the corresponding one-loop counterterm
a most careful analysis is required, because the 3-point and 4-point contributions cannot be simply absorbed by a
renormalization of the YM coupling constant $g$, but also need a field renormalization
\begin{equation}
A_{r,\mu} = \left(1+c_{0}\frac{C_{2}(G)}{32\pi^{2}\epsilon}\right)A_{\mu},
\end{equation}
{which absorbs all $\alpha$-dependence of $c_0$. On the other hand the $\alpha$-independence of one-loop radiative corrections to the coupling constant $g$ follows from the BRST invariance of the theory.}
Moreover, as it is well known in the last case the renormalization    process    has to be extended to any loop order.

The effect of higher derivative terms leads to a modified $\beta$-function.
In the case $n\geqslant 1$, the result is
\begin{equation}
\beta_n= c_n{
{\frac{g^3 C_{2}(G)}{32 \pi^{2}}}},
\end{equation}
whereas in the case $n=0$, once the wave-functions of gluons and ghosts are renormalized, one gets the standard Yang-Mills  $\beta$-function
of the coupling constant \cite{af2}.

Summing up all orders of perturbation theory one gets the renormalization group flow of the bare coupling constant 
\begin{equation}
g_{_{\mathrm{bare}}}^2(\mu)= \frac{g^2}{1 - \frac{g^2 C_{2}(G)}{(4 \pi)^2} c_n 
\log \mu/\Lambda{
}},
\end{equation}
which is running to $0$ as the renormalization scale $\mu$ goes to infinity.
Since the higher derivative terms of the  action  (\ref{hd}) do not get any divergent radiative correction,  the coefficient 
$g^2 \Lambda^{2n}=g_{_\mathrm{bare}}^2 \Lambda_{_\mathrm{bare}}^{2n}$ remains unrenormalized. Thus,
the ultraviolet behavior of  the bare mass $\Lambda_{_{\mathrm{bare}}}$  parameter is just  the oposite of $g_{_{\mathrm {bare}}}$, i.e. $\beta_\Lambda=-\frac{\Lambda }{n g} \beta_n$, which implies that $\Lambda_{_{\mathrm{bare}}}$ is running to infinity as the renormalization scale $\mu$ goes to infinity. 

{
This observation allows to extend the family of consistent higher derivative theories beyond the expected tree-level bounds, because the masses of  all pathological ghosts breaking causality and unitarity are proportional to $\Lambda_{_{\rm{bare}}}$, and  for asymptotically free theories these ghosts  become infinitely heavy and decouple from the physical spectrum in the UV regime. }

\section{Asymptotic Freedom}

The behavior of the renormalized two-point  functions  is more involved. The one-loop result at leading order  in
$\Lambda$ is
\begin{equation}
\Gamma_{\mu\nu}^{ab}(p)=- \frac{C_{2}(G)}{   32\pi^{2}}i\delta^{ab}\left(p^{2}\eta_{\mu\nu}-p_{\mu}p_{\nu}\right) \Pi(p^2),
\end{equation}
with
\begin{equation}
\Pi(p^2)=\left ( b_n\log\frac{p^2+\Lambda^2}{\Lambda^2}   +    c_0 \log\frac{p^2}{{\Lambda^2_{\rm QCD}}} \right),
\label{pidef}
\end{equation}
where we identified an arbitrary renormalization scale $\mu$ with the higher derivative scale $\Lambda$
and
\begin{equation}
 b_0 =0,   \quad  b_1=-10-\alpha, \quad   \hbox{and} \quad  b_n=14-\alpha-23n+5n^{2}\
 \quad   \hbox{for}\  n \geqslant 2 
\label{one}
\end{equation}
 or generally $b_n=c_n-c_0$ for any $n$.

{} There are two different asymptotic regimes, the $p\gg \Lambda$ UV  regime, where
\begin{equation}
\Gamma_{\mu\nu}^{ab}(p)= - c_n\frac{C_{2}(G)}{   32\pi^{2}}i\delta^{ab}\log\frac{p^2}{\Lambda^2}\left(p^{2}\eta_{\mu\nu}-p_{\mu}p_{\nu}\right)
\end{equation}
and  the  intermediate IR regime $  \Lambda_{\rm QCD}  < p\ll \Lambda$, where   we recover the renormalized two-point function of  standard  QCD 
\begin{equation}
\Gamma_{\mu\nu}^{ab}(p)=   - c_0   \frac{C_{2}(G)}{   32\pi^{2}}i\delta^{ab}\log\frac{p^2}{\Lambda^2_{\rm QCD}}\left(p^{2}\eta_{\mu\nu}-p_{\mu}p_{\nu}\right). 
\end{equation}

\begin{figure}[h!]
\centerline{\includegraphics[width=15cm]{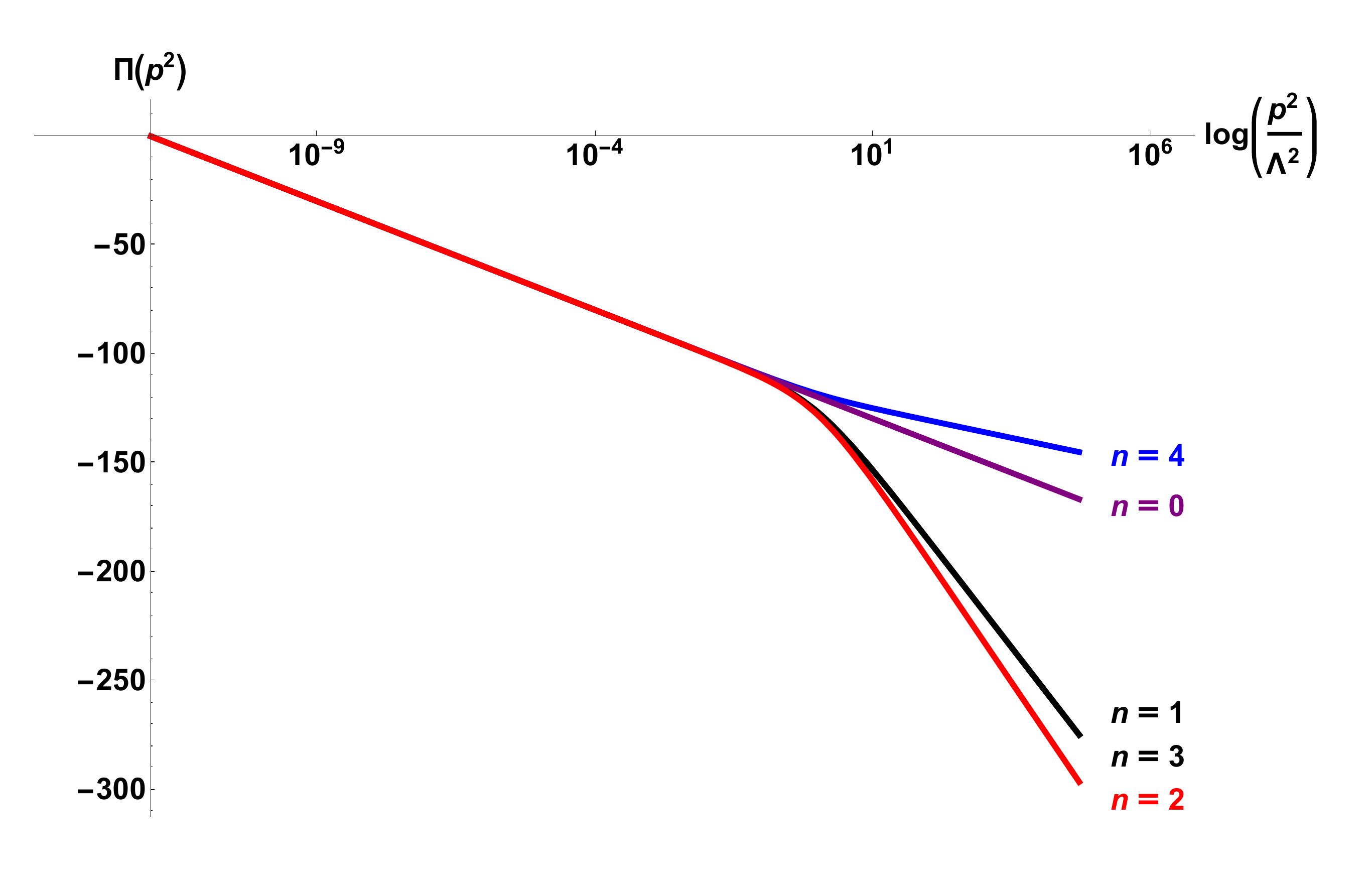}}\caption{
Momentum dependence of  one-loop radiative corrections to  the 2-point function form factor $\Pi(p^2)$ for different higher derivative theories. The cases $n=1$ and $n=3$ give rise to the same form factors.    The results for large momenta are independent of the gauge fixing parameter $\alpha$, whereas in the IR we used a  Lorentz  gauge with $\alpha=0$.
    In the infrared regime  $\Lambda_{\rm QCD}<p\ll \Lambda$,
the  behavior is independent on the number of extra derivatives and agrees with that of standard gauge theories. In the UV regime $p\gg\Lambda$,  the behavior depends on the number of extra derivatives, but in  all five cases the theory is asymptotically free. We assumed that  the higher derivative scale $\Lambda$ is of the order of $100$\,TeV, i.e. $\Lambda  \approx  10^6 \Lambda_{\rm QCD}\,.$ }
\label{fig 2020}
\end{figure}

The renormalization group (RG) flow of the higher derivative Yang-Mills theory interpolates for $n\geqslant 1$ between
these two asymptotic regimes with two different beta functions 
\begin{equation}
\beta_{\rm UV}=
c_n{
{\frac{g^3 C_{2}(G)}{ 32 \pi^{2}}}} \quad{\rm and}\quad \beta_{\rm IR }=-\frac{22}{3}
{\frac{g^3 C_{2}(G)}{32\pi^{2}}}.
 \end{equation}
In the infrared regime one recovers the standard values of    the $\beta$-function for    asymptotically free Yang-Mills  scenario.
However, in the ultraviolet regime there is a variety of scenarios involving either asymptotic freedom or asymptotic
slavery, {except in the case $n=0$ where the $\beta$-function remains frozen in the QCD asymptotically free regime for all $p> {\Lambda_{\rm QCD}}$ (see Figure \ref{fig 2020}).}

\noindent

\medskip

Assuming that $n$ has
to be an integer in order to preserve locality, the number of UV asymptotically free  theories is very limited. It reduces to theories with negative  coefficients of
the $\beta$-function,  i.e.,  only to five cases $n=0, 1,2,3,4$: 
\begin{equation}
\widetilde{c}_0=-\frac{22}{3}\,,\quad  c_1=-\frac{43}{3}\,,\quad c_2=-\frac{49}{3}\,,\quad c_3=-\frac{43}{3}\,, \quad c_4=-\frac7{3} \,. 
\end{equation}
For all other    integer    values of $n$, $c_n>0$.
This shows how the limitation to UV asymptotically free theories imposes severe constraints
on higher derivative gauge theories.

In order to enlarge the family of theories with UV asymptotically free regimes one can introduce an extra
dimensionless coupling    $\lambda$    in the theory, and  replace the Hodge Laplacian operator (\ref{HL})  in (\ref{hd}) by  the
operator $^\lambda\Delta$ defined by
\begin{equation}
^\lambda\Delta{}_{\mu\, a}^{\nu\, b}=    -\delta^b_a\delta^\nu_\mu D^2   +2 \lambda f^b{}_{ca}
\, F_{\mu}{}^{\nu c}\,.
\label{extLaplacian}
\end{equation}
For  $\lambda\neq1$  instantons ($d_A^\ast  F =0$) are not anymore solutions of the Euclidean equations of motion.
The Euclidean extension of  operator $^\lambda\Delta$ is positive for
$\lambda = 1$ and  $\lambda= 0$, but not for  $n$
 odd and larger than $1$. However, the action of the theory remains positive for any even $n$,
 which is a physically relevant requirement.
In this case, the structure of one-loop divergences gets modified because now 
\begin{eqnarray}
 c_n= \begin{cases} \displaystyle
      -9\lambda^{2}
      -18\lambda{}+
      \frac{38}{3}
, &  \hbox{for}\  n=1\nonumber
\\[3ex]
 \displaystyle
 \left(5n^{2}-18n+16\right)\lambda^2
      {{-}}\left(4n^{2}+10n+4\right)\lambda+
      4n^{2}+5n-\frac{7}{3}
 , &  \hbox{for}\  n\geqslant 2.\hfill
\nonumber
\end{cases}
\end{eqnarray}
Although  these coefficients reduce to (\ref{ccoeff})  in a continuous way when $\lambda \to1$.

   The coupling $\lambda$ does not  get any divergent  radiative contribution and, thus, is kept fixed under RG flow as an extra  parameter of the theory.
On the other hand, for higher values of $\lambda\gg 1$ the constraint of having a window with asymptotic freedom is even more stringent.
   In fact, in the limit $\lambda\to +\infty$, only the theory with $n=2$ is marginally asymptotically free. Actually, for $n=2$ and $\lambda\to+\infty$ the theory becomes UV-finite.
Notice that the minimal   choice    of   scalar covariant Laplacian $^0\Delta=-D^2$   never gives rise to asymptotically free theories.
In fact, the above results imply that the  Hodge-covariant Laplacian $\Delta$ is the optimal choice to get  asymptotic freedom in a  larger number of higher derivative gauge theories.

\vspace{-.0cm}

\begin{figure}[h!]
\begin{center}
\hspace{.5cm}{\includegraphics[width=13cm]{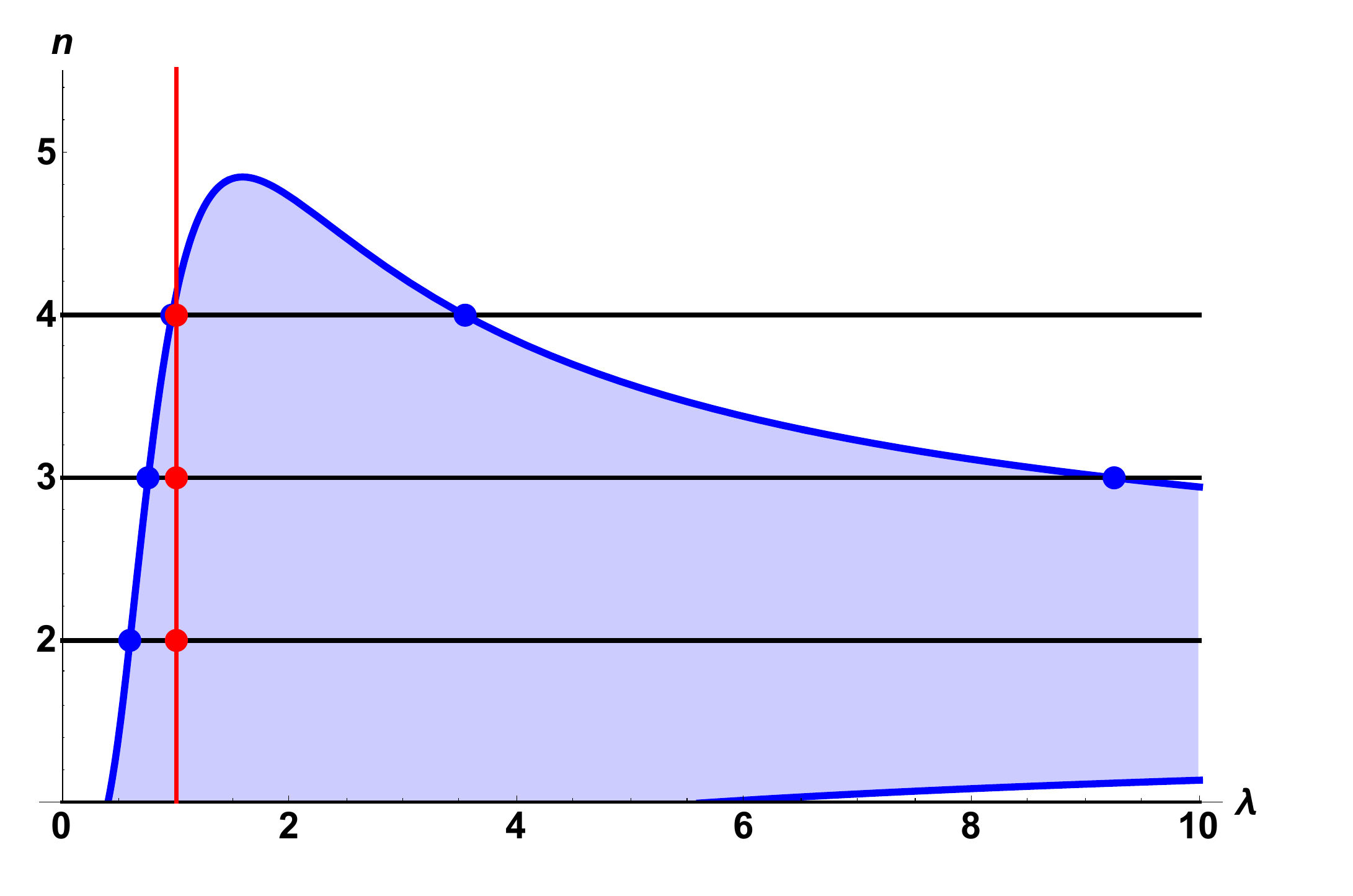}}
\end{center}

\caption{
Domain of asymptotically free gauge theories (shadow blue area). 
Horizontal lines correspond to local theories and $\lambda$
is the extra parameter of the generalized Laplacian operator (\ref{extLaplacian}).
Red points mark  theories with Hodge Laplacians ($\lambda=1$)
and blue points  UV-finite theories. In the limit $\lambda\to \infty$ the upper and lower curves converge at $n=2$.
}
\label{fig 2021}
\end{figure}
\noindent
\vspace{-.8cm}

It is also interesting to remark that the theories  are UV-finite for special values of the parameter $\lambda$:
\vspace{-.2cm}
\begin{center}
 \begin{tabular}{c  c c }
$ n=1$ &$ \lambda_1= -2.55$ & $\lambda_2=0.55$ \\
 $n=2$ &$\lambda_{\phantom{1}}=\phantom{-}0.59$  
 \\
 $n=3$ & $\lambda_1= \phantom{-}0.75$& $\lambda_2=9.25$  \\
 $n=4$ & $\lambda_1=\phantom{-}0.96$ & $\lambda_2=3.54$
\end{tabular}
\end{center}
without any need of supersymmetry.  Notice that the introduction of operator ${}^\lambda\Delta$ recalls the addition of terms cubic or quartic in field strengths to the action of a higher derivative theory, which is crucial for achieving UV-finiteness \cite{MR, MR2,fingauge}.

For $G=SU(N)$ the interaction with fundamental quarks generates
a lower   (absolute)     value of the $\beta$-function 
 \begin{equation}
\beta_n=\left(c_n N+ {\textstyle 
  \frac{ 4}{3}   N_q}\right)
{\frac{g^3 }{32\pi^{2}}}\, %
\end{equation}
 for $n\geqslant1$, and  thus,
 slightly  stronger constraints    that shrink    
   the window of UV asymptotically free regime.
In the case $\lambda=1$, the window is preserved for small number of quarks $N_q< \frac{43}{4} N$. For a larger number of quarks the window shrinks till its disappearance in the large $N_q$ limit.

\section{Discussion}

In summary, gauge theories with higher derivatives  can be asymptotically free
in the UV regime, if the    total number of higher derivatives    is not bigger than 8. Otherwise, the theories
become UV-slave and exhibit a strong interacting regime both in the UV and in the
deep IR. There is an intermediate regime $\Lambda^2_{\rm QCD}< p^2 \ll \Lambda^2$, where the theory
is always asymptotically free with the same  scaling properties as in the  standard
QCD  at high energies.

This means that there are UV-completions of gauge theories, where  the behavior of the theory for energy scales larger than $\Lambda$ can dramatically change from asymptotic freedom to asymptotic slavery.

{
One remarkable feature of higher derivative theories is the possibility of having an UV scale-invariant fixed point
with $\beta=0$, where the theory becomes finite,    without any UV divergence \cite{fingauge}.    The behavior of these theories in the presence of
a $\theta$-term deserves further analysis. In particular, it will be interesting to find the appearance of non-standard
effective potentials,  which might have implications for  axion physics phenomenology.}

An  interesting remaining open problem is the analysis of the dynamics of the hidden ghost sector   and its implications for
 unitarity, causality and stability of these theories. Ghost fields appear in conjugate
 pairs of poles of the propagator  as in the case of Lee-Wick  theories \cite{leewick}--\cite{AP2}.
 However,  in the present case  of asymptotically free theories the running of the bare mass $\Lambda_{_{\mathrm{bare}}}$ towards  $+\infty$ in the UV regime can make the pathological effects of ghosts  harmless. It is remarkable that this property only holds for asymptotically free theories, which enhances the relevance of such a UV behavior  for the consistency of higher derivative theories.
 
 These results are compatible with the  behavior  of lattice gauge theories. The Wilson formulation of lattice gauge theories satisfies in the Euclidean formalism the reflection positivity property which guarantees  unitarity and stability of the corresponding quantum theory. The first two leading terms in the continuum limit  of Wilson's action of lattice gauge theories do coincide with those of (\ref{hd}) for $n=1$. On the other hand, non-abelian lattice gauge theories turn out to be also asymptotically free. These two relevant properties can provide an ultimate argument for the consistency of 
 asymptotically free higher derivative gauge theories.
 
 In spite of this fact some physical effects of the pairs of complex poles can still be traced down in
the behavior of the trace anomaly on curved backgrounds. Under  certain consistency conditions $a$-theorem establishes
  that  the coefficient  of the Gauss-Bonnet term of  the anomaly $a$ must evolve in a monotonically decreasing way under the  renormalization group flow     \cite{atheorem}.    However, in the case of the UV asymptotically free theories analyzed in this  paper  $a$  is positive in standard infrared regime whereas in the UV regime it can reach negative values \cite{PS,GH,ARSh}. This breaking of $a$-theorem raises some questions that require a deeper analysis. 
 
The above results open a new interesting perspective for the analysis of higher derivative theories of  quantum
gravity, which deserves further study.

\section*{Acknowledgements}
The work of M.A. and F.F. is partially supported by Spanish MINECO/FEDER
grant  PGC2018-095328-B-I00 and DGA-FSE grant 2020-E21-17R, and COST
action programs  MP1405-37241 and QGMM-CA18108. The work of L.R. was partially supported by
a  Short Term Scientific Mission (STSM), within the COST action
 MP1405-37241 for the group Effective Theories of Quantum Gravity
 and Quantum Structure of Spacetime WG3, in QSpace program. L.R. would
 like to thank Department of Theoretical Physics, 
 of University of Zaragoza for the kind hospitality during the initial stage of this project.
\vfill
\appendix{}
\section{Two-point functions: One-loop  gluonic UV divergences. }

\begin{figure}[h!]
\begin{center}
\hspace{.1cm}{\includegraphics[width=13cm]{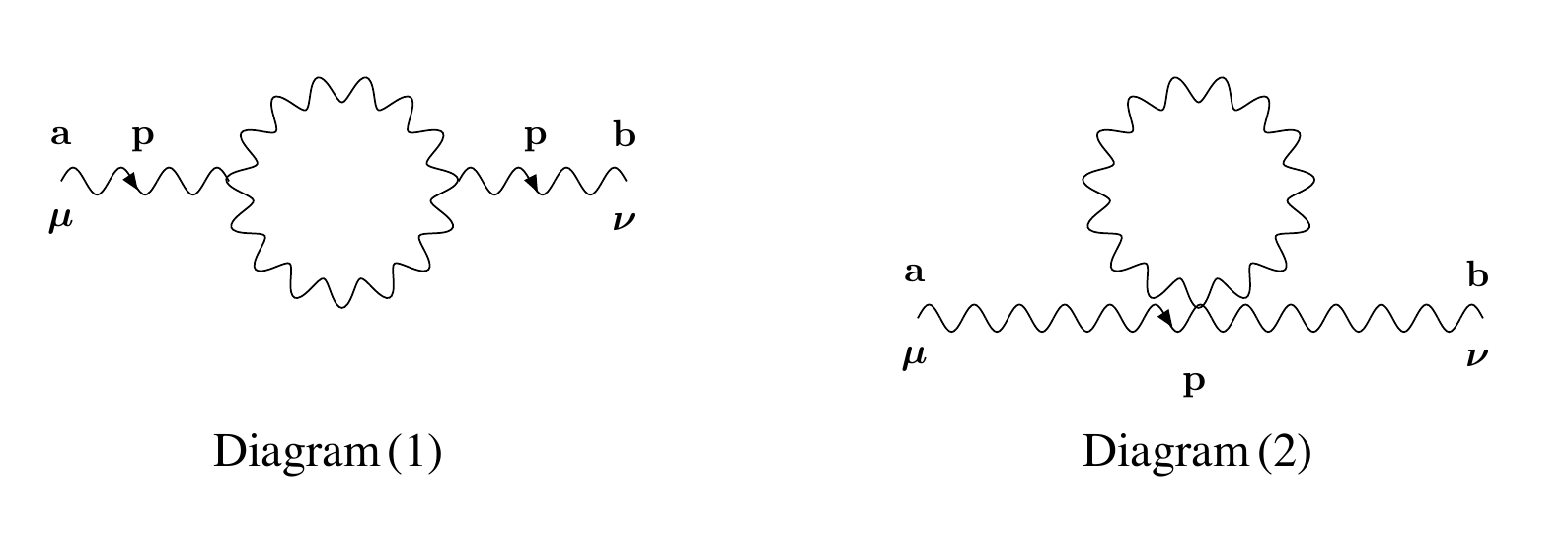}}
\end{center}
\vspace{-.6cm}
\caption{
Gluonic one-loop  Feynman diagrams contributing to the 2-point functions of higher derivative theory.}
\label{fig 2023}
\end{figure}

One-loop  pure  gluonic  contribution 
to the two-point gluon function is given by Feynman diagrams of Figure\, \ref{fig 2023}.

For theories with  $n>1$ the UV-divergent contribution   of  diagram (1) of Figure\,\ref{fig 2023} is given in dimensional regularization by
\begin{equation}
\begin{aligned}
  ^{(1)}\Gamma^{ab}_{\mu\nu}(p)&=\frac{iC_2(G)\delta^{ab}}
  {64\pi^{2}\epsilon}p^{2}\eta_{\mu\nu}\displaystyle\Bigl[\left(4n^{2}+16n+16\right)\lambda^{2}+\left(8n^{2}+32n+32\right)\lambda
\\
&\hskip 1.5cm+\frac{2n^{4}}{3}+\frac{8n^{3}}{3}-8n^{2}+\frac{8n}{3}+\frac{50}{3}
+\left(\frac{10n^{2}}{3}+\frac{32n}{3}+8\right)\alpha\Bigr]\\
&
-\frac{iC_2(G)\delta^{ab}}{64\pi^{2}\epsilon}p_{\mu}p_{\nu}
\Bigl[\left(4n^{2}+16n+16\right)\lambda^{2} 
+\left(8n^{2}+32n+32\right)\lambda \\
&\hskip 1.3cm-\frac{4n^{4}}{3}-\frac{16n^{3}}{3}-16n^{2}+\frac{8n}{3}+\frac{56}{3}+\left(\frac{4n^{2}}{3}+\frac{20n}{3}+8\right)\alpha
\Bigr],
 \label{diagrama}
 \end{aligned}
\end{equation}
where  $\epsilon=4-d$.

The  corresponding  UV-divergent contribution  of diagram (2) of Figure\,\ref{fig 2023} is
\begin{equation}
\begin{aligned}
^{(2)}\Gamma^{ab}_{\mu\nu}(p)&
= \frac{iC_2(G)\delta^{ab}}{64\pi^{2}\epsilon}p^{2}\eta_{\mu\nu}
\displaystyle\Bigl[\left(-24n^{2}+56n-80\right)\lambda^{2}+\left(8n^{2}+8n-16\right)\lambda
\\
&\hskip 2.cm- \frac{2n^{4}}{3}-\frac{8n^{3}}{3}-8n^{2}-\frac{68n}{3}-8
-\left(\frac{10n^{2}}{3}+\frac{32n}{3}+8\right)\alpha\Bigr]
\\
&-\frac{iC_2(G)\delta^{ab}}{64\pi^{2}\epsilon}p_{\mu}p_{\nu}
\Bigl[\left(-24n^{2}+56n-80\right)\lambda^{2} 
+\left(8n^{2}+8n-16\right)\lambda \\
&\hskip 3cm
+\frac{4n^{4}}{3}+\frac{16n^{3}}{3}-\frac{68n}{3}-8
-\left(\frac{4n^{2}}{3}+\frac{20n}{3}+8\right)\alpha\Bigr].
 \label{diagramb}
 \end{aligned}
\end{equation}
The sum of these diagrams 
\begin{equation}
\begin{aligned}
\!\!\!\!\!\!  ^{(1)}\Gamma^{ab}_{\mu\nu}(p)+\, ^{(2)}\Gamma^{ab}_{\mu\nu}(p)&
  =-\frac{iC_2(G)\delta^{ab}}{16\pi^{2}\epsilon}
 \displaystyle\Bigl[
   \left(5n^{2}-18n+16\right)\lambda^{2}
   -\left(4n^{2}+10n+4\right)\lambda\\
   &\hskip 0.5cm
   +\, 4n^{2}+5n-\frac{13}{6}\Bigr]\left(p^{2}\eta_{\mu\nu}-p_{\mu}p_{\nu}\right)
- \frac{iC_2(G)\delta^{ab}}{32\pi^{2}\epsilon}\ p_{\mu}p_{\nu}
 \label{diagram}
 \end{aligned}
\end{equation}
cancels out the $n^{4}$ and $n^{3}$ terms and the $\alpha$-dependence. The cancelation of $\alpha$-dependence can be understood in  background gauge formalism \cite{alsh,BV} as a consequence of the fact 
that the infinitesimal variation of the effective action under changes of $\alpha$ is given by a functional equation where one of the factors is proportional to the equations of motion of the theory \cite{alsh}. Since these equations involve higher derivative terms which do no get any divergent contribution from radiative corrections we can conclude that there are no $\alpha$-dependent  UV divergences due to one-loop radiative corrections to any  gluonic  $n$-point function, which is in agreement with the above explicit calculation (\ref{diagram}).

\begin{figure}[h!]
\begin{center}
\hspace{.1cm}{\includegraphics[width=8cm]{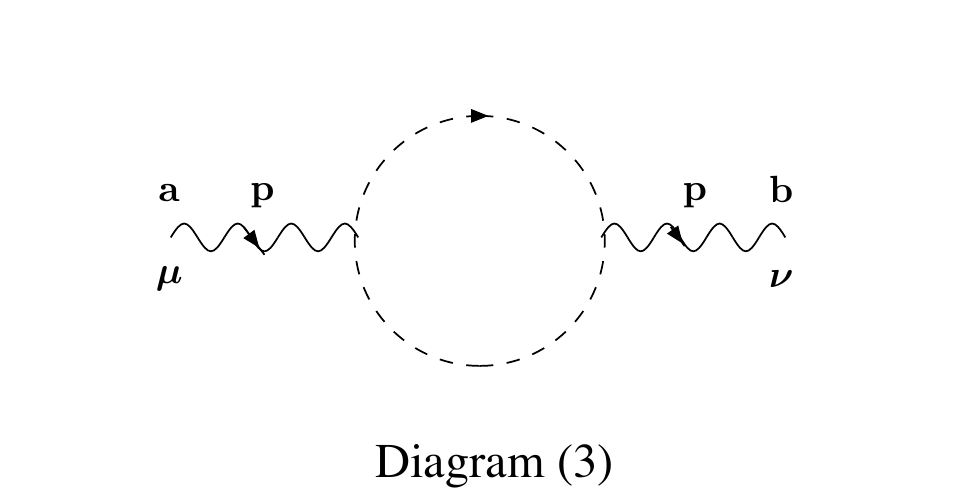}}
\end{center}
\vspace{-.6cm}
\caption{
One-ghost-loop Feynman diagram contributing to the 2-point functions.}
\label{fig 2024}
\end{figure}

The contribution of Faddeev-Popov ghosts to $\Gamma^{ab}_{\mu\nu}$  is given by diagram (3) of Figure\,\ref{fig 2024}
\begin{equation}\label{ghost}
^{\mathrm{FP}}\Gamma^{ab}_{\mu\nu}(p)=\frac{i C_2(G)\delta^{ab}}{32\pi^{2}\epsilon}\left[\frac13(p^{2}\eta_{\mu\nu}-p_{\mu}p_{\nu})+p_{\mu}p_{\nu}\right].
\end{equation}

Notice that it is the same as that of the standard Faddeev-Popov ghosts 
in ordinary Yang-Mills theory. This is a consequence 
of the factorization of the higher derivative Faddeev-Popov determinant 
\begin{equation} 
\det [(-\partial^\sigma \partial_\sigma)^{n} (-\partial^\mu D_\mu )]= \det (-\partial^\sigma \partial_\sigma)^{n}  \det ( -\partial^\mu D_\mu)
\end{equation}
and the fact that $-(\partial^\sigma \partial_\sigma)^{n} $ does not give any contribution to the gluonic $n$-point functions.

Finally, the sum of all diagrams contributing to the two-point function
reads
\begin{equation}
\begin{aligned}
\Gamma^{ab}_{\mu\nu}(p)&=
 -\frac{iC_2(G)\delta^{ab}}{16\pi^{2}\epsilon}
 \displaystyle\Bigl[
   \left(5n^{2}-18n+16\right)\lambda^{2}
   -\left(4n^{2}+10n+4\right)\lambda\\
   &\hskip 3cm
   +\, 4n^{2}+5n-\frac{7}{3}\Bigr]\left(p^{2}\eta_{\mu\nu}-p_{\mu}p_{\nu}\right).
 \label{total}
 \end{aligned}
\end{equation}
The explicit transverse structure of the two-point function $\Gamma^{ab}_{\mu\nu}(p)$ is a consequence of BRST invariance of 
the theory. 

Diagrams involving two external ghosts and any number of external gluons are finite because in the
corresponding Feynman graphs there is always one more gluon propagator with higher derivatives than vertices
with higher derivatives. This implies that the gluonic wave function does not require renormalization. Thus all divergences can be absorbed by gauge coupling renormalization that by BRST symmetry cannot depend on the gauge fixing parameters. This fact explains why in higher derivatives theories the gluonic 2-point function does not depend on 
the gauge fixing parameter $\alpha$ as shown by the explicit calculation  (\ref{total}). Notice that for this reason the result
holds even for gauge fixing conditions with less derivatives whenever the number of extra derivatives of the gauge fixing terms is larger than one. This fact guarantees that both the transverse and longitudinal parts of the wave functions do not
acquire any divergent contribution from radiative corrections.

Similar results hold for the UV-divergent contributions in $n=1$ higher derivative theories. The gluonic  one-loop  contribution to $\Gamma^{ab}_{\mu\nu}(p)$ is  given by the two diagrams of Figure \ref{fig 2023}. The contribution of 
diagram (1) of Figure \ref{fig 2023} is
\begin{equation}
\begin{aligned}
  ^{(1)}\Gamma^{ab}_{\mu\nu}(p)&=\frac{iC_2(G)\delta^{ab}}{32\pi^{2}\epsilon}\displaystyle\left[\left(18 \lambda^{2}+36\lambda+\frac{22}{3}+11 \alpha\right)
    (p^{2}\eta_{\mu\nu}-p_{\mu}p_{\nu}) +
\left(8+3\alpha\right)p_{\mu}p_{\nu}\right]
 \label{diagramaa}
 \nonumber
 \end{aligned}
\end{equation}
and that of diagram (2) is
\begin{equation}
\begin{aligned}
^{(2)}\Gamma^{ab}_{\mu\nu}(p)&=-\frac{iC_2(G)\delta^{ab}}{32\pi^{2}\epsilon}\displaystyle\left[\left(33+11 \alpha\right)(p^{2}\eta_{\mu\nu}-p_{\mu}p_{\nu}) +
\left(9+3\alpha\right)p_{\mu}p_{\nu}\right].
 \label{diagrambb}
 \end{aligned}
\end{equation}
And in the sum of these two diagrams, \begin{equation}
\begin{aligned}
  \Gamma^{ab}_{\mu\nu}(p)&=\frac{iC_2(G)\delta^{ab}}{32\pi^{2}\epsilon}
  \displaystyle\left[\left(18\lambda^2 +36 \lambda-\frac{77}{3}\right)(p^{2}\eta_{\mu\nu}-p_{\mu}p_{\nu}) - p_{\mu}p_{\nu}\right]
 \label{diagramab}
 \end{aligned}
\end{equation}
the $\alpha$-dependence cancels out by the same reasons that in the $n\geqslant2$ case.

When we add the contribution  of  diagram (3) of Figure \ref{fig 2024}
with one FP ghost loop given in (\ref{ghost})
we finally obtain
\begin{equation}
\begin{aligned}
  \Gamma^{ab}_{\mu\nu}(p)&=\frac{iC_2(G)\delta^{ab}}{16\pi^{2}\epsilon}\left(9\lambda^{2}+18\lambda-\frac{38}3\right)\left(p^{2}\eta_{\mu\nu}-p_{\mu}p_{\nu}\right),
 \end{aligned}
\end{equation}
which again has a transverse structure in agreement with BRST symmetry.

\end{document}